\documentclass{article}
\usepackage{spconf,amsmath,graphicx,bm,amsfonts,mathrsfs,amssymb,subfigure,pifont}
\usepackage{algpseudocode}
\usepackage[Algorithmus]{algorithm}
\usepackage{cite,url}
\usepackage{hyperref}

\usepackage{tikz}
\usetikzlibrary{bayesnet}
\usetikzlibrary{arrows,shapes,snakes,automata,backgrounds,petri}

\allowdisplaybreaks

\definecolor{00BCB4}{RGB}{0,188,180}
\definecolor{C4E86B}{RGB}{196,232,107}
\definecolor{49DEB2}{RGB}{73,222,178}
\definecolor{FF4747}{RGB}{255,71,71}
\definecolor{FF3561}{RGB}{255,53,97}
\definecolor{CD2C6C}{RGB}{205,44,108}
\definecolor{10B48E}{RGB}{16,180,142}
\definecolor{8FDBD8}{RGB}{143,219,216}

\newcommand{\vx}{{\bm x}}
\newcommand{\vy}{{\bm y}}
\newcommand{\vz}{{\bm z}}

\newcommand{\vr}{{\bm r}}
\newcommand{\vs}{{\bm s}}
\newcommand{\vw}{{\bm w}}

\newcommand{\mH}{{\bm H}}
\newcommand{\mX}{{\bm X}}
\newcommand{\mA}{{\bm A}}
\newcommand{\mB}{{\bm B}}

\newenvironment{signal_prior}[1][\ding{113} Sparse signal model:\\]{\begin{trivlist}\item[\hskip \labelsep {\bfseries #1}]}{\end{trivlist}}
\newenvironment{noise_prior}[1][\ding{113} Quantization noise model:\\]{\begin{trivlist}\item[\hskip \labelsep {\bfseries #1}]}{\end{trivlist}}

\title{Bayesian Massive MIMO Channel Estimation \\with Parameter Estimation using Low-Resolution ADCs}
\name{Shuai Huang$^*$, Deqiang Qiu$^*$ and Trac D. Tran$^\dagger$}
\address{$^*$Emory University, Atlanta, GA 30322, USA\\ $^\dagger$Johns Hopkins University, Baltimore, MD 21218, USA}
\begin{document}
\ninept
\maketitle
\begin{abstract}
In order to reduce hardware complexity and power consumption, massive multiple-input multiple-output (MIMO) systems employ low-resolution analog-to-digital converters (ADCs) to acquire quantized measurements $\vy$. This poses new challenges to the channel estimation problem, and the sparse prior on the channel coefficient vector $\vx$ in the angle domain is often used to compensate for the information lost during quantization. By interpreting the sparse prior from a Bayesian perspective, we can assume $\vx$ follows certain sparse prior distribution and recover it using approximate message passing (AMP). However, the distribution parameters are unknown in practice and need to be estimated. Due to the increased computational complexity in the quantization noise model, previous works either use an approximated noise model or manually tune the noise distribution parameters. In this paper, we treat both signals and parameters as random variables and recover them jointly within the AMP framework. The proposed approach leads to a much simpler parameter estimation method, allowing us to work with the quantization noise model directly. Experimental results show that the proposed approach achieves state-of-the-art performance under various noise levels and does not require parameter tuning, making it a practical and maintenance-free approach for channel estimation.
\end{abstract}
\section{Introduction}
\label{sec:intro}
Massive multiple-input multiple-output (MIMO) technology has shown great potential in the next generation wireless systems such as 5G cellular network \cite{Larsson:massiveMIMO:2014,Lu:massiveMIMO:2014,Asaad:massiveMIM0:2018}. It deploys large numbers of antennas (in the order of hundreds or more) at the base station to serve multiple users simultaneously within the same physical environment. Massive MIMO not only produces high throughput but also boosts the array gain so that the noise and interference can be greatly reduced. On the other hand, large antenna arrays increase the complexity of hardware design and the power consumption drastically. For example, the power consumed by analog-to-digital converters (ADCs) in the hardware grows exponentially with the number of quantization bits \cite{Walden:ADC:1999}. Currently it is either too expensive or impractical to use high-resolution ADCs in base stations and portable devices \cite{Murmann:1997-2020}. As a result, there has been a growing interest in low-resolution ADCs that output $1\sim3$ bits in recent years. In particular, 1-bit ADC does not require automatic gain control or linear amplifier, and is much preferred in wideband millimeter wave (mmWave) communication systems that require high sampling frequency \cite{Rappaport:MmWave:2015}.

In order to achieve reliable data transmission and maximize channel capacity in this challenging setting, we need to estimate the channel state information so that we can compensate for common distortion effects such as scattering, fading, and power decay. The nonlinear (quantized) measurements from low-resolution ADCs pose new challenges to channel estimation. Dithering has been used in \cite{Dabeer:CEADC:2010,Zeitler:DitherCE:2012} to alleviate the nonlinearity of the quantization noise. However, it becomes less effective for MIMO channel where signals from multiple antennas are linearly combined before quantization. Fortunately, the ``\emph{channel sparsity}'' of the MIMO system can be exploited to improve the channel estimation performance from quantized measurements. In massive MIMO channels, the channel energy is mostly concentrated in small regions of the multipath angular spread, and the channel matrix is thus approximately sparse in the angle domain \cite{Sayeed:MIMOChannel:2002,Rao:CESparse:2014}. In particular, the channel matrix in mmWave MIMO channel is approximately sparse in both the angle domain and delay domain \cite{5G:Report:2016}. Compressive sensing techniques then enable us to compensate for the information lost during quantization by incorporating this sparse prior into the channel estimation problem \cite{Decode05,CS06}.

Various sparse signal recovery methods can be used to estimate the channel matrix from quantized measurements, and they generally differ in how the sparse prior is enforced. For example, iterative hard thresholding (IHT) \cite{Blumensath2008,Jacques:1bitCS:2013} imposes a direct constraint on the sparsity of the signal $\vx$, i.e. $\|\vx\|_0\leq K$, where $K$ is the number of nonzero entries in $\vx$, whereas others promote sparse solutions through regularization functions like the $l_1$-norm $\|\vx\|_1$ \cite{l1stable06}, the $l_p$-norm $\|\vx\|_p$ \cite{Nonconvex_lp07}, and generalized entropy functions \cite{Huang:Entropy:2019}.

The sparse prior can also be interpreted from a Bayesian perspective. In this approach, the signal $\vx$ is assumed to follow some sparse prior distributions $p(\vx|\boldsymbol\lambda)$ like Laplace distribution and Bernoulli-Gaussian mixture distribution, where $\boldsymbol\lambda$ contains the distribution parameters. Let $\vy$ denote the quantized measurements. The posterior distribution $p(\vx|\vy)$ can be computed via approximate message passing (AMP) \cite{Minka:2001}, and the minimum mean squared error (MMSE) estimation of the signal is then $\hat{\vx}=\mathbb{E}[\vx|\vy]$. Both the denoising formulation \cite{Donoho:AMP:2009,Metzler:Denoising:2016,Ma:AMP_Denoise:2016} and the Bayesian formulation \cite{Rangan:GAMP:2011,Krzakala:2012:1} of AMP can be used to recover the sparse signal $\vx$. AMP has been used for channel estimation in massive MIMO channels due to its computational efficiency and superior performance \cite{Wen:LowADC:2016,Mo:LowADC:2018,Bellili:Lap:2019}.

In practice, the distribution parameters in AMP are \emph{unknown} and need to be estimated. Previous works relied on expectation maximization (EM) \cite{Dempster:EM:1977} to find the parameters that maximize the likelihood \cite{Vila:EMGM:2013}. However, the involved computation becomes increasingly difficult in the case of complicated probability models such as the quantization noise model of low-resolution ADCs. As a result, the simple additive white Gaussian noise (AWGN) model was often adopted to approximate the quantization noise \cite{risi2014massive, Wang:Multiuser:2015,Wen:Channel:2015}, leading to suboptimal performance. The approach in \cite{Mo:LowADC:2018} employed the actual quantization noise model and only estimated the signal distribution parameters, whereas the noise distribution parameter was prespecified or manually tuned for different types of channels, which greatly limited its applicability in real-life applications.

By treating the distribution parameters as random variables, we proposed an extension to the AMP framework in \cite{PE_GAMP17} where the posteriors of both signal and parameters can be computed and used to recover them jointly. We later present a more computationally efficient approach to perform parameter estimation in \cite{Huang:1bitCS:2020} and show that our proposed approach has a wider applicability and could directly work with the complicated quantization noise model. In this paper, we apply the proposed approach to solving the particular channel estimation problem in broadband mmWave MIMO channel. Compared to previous AMP-based channel estimation methods that either use an approximated noise model or manually tune the noise distribution parameter, the proposed approach estimates the channel matrix based on the actual quantization noise model and comes with ``\emph{built-in}'' parameter estimation. 

This paper proceeds as follows. In Section \ref{sec:problem_formulation}, we introduce the sparse channel estimation problem for the single-user case in the angle domain and the probability models of the signal and noise. Section \ref{sec:amp_pe} discusses the AMP framework with built-in parameter estimation and how it can be effective in computing the MMSE estimation of the channel coefficients. Section \ref{sec:experiments} presents our experimental results with careful bench-marking comparison to other state-of-the-art methods. Finally, we offer concluding remarks in Section \ref{sec:conlucsion}.

\section{Problem Formulation}
\label{sec:problem_formulation}
When the channel matrix $\mH$ represents the baseband channel impulse response from the transmitter to the receiver in the ``antenna domain'', it is typically dense. If $\mH$ is transformed into the ``angle domain'' \cite{Sayeed:MIMOChannel:2002}, the transformed channel coefficients in $\mX$ become approximately sparse. We next briefly introduce the different domain representations of the channel in a single-user MIMO system according to the derivations outlined in \cite{Mo:LowADC:2018}.

Suppose there are $N_t$ antennas at the transmitter and $N_r$ antennas at the receiver, and the delay spread is limited to $L$ intervals. The quantized measurements $\vy[k]\in\mathbb{C}^{N_r}$ at time $k$ are
\begin{align}
    \vy[k]=\mathcal{Q}\left(\sum_{l=0}^{L-1}\mH[l]\vs[k-l]+\vw[k]\right)\,,
\end{align}
where $\mH[l]\in\mathbb{C}^{N_r\times N_t}$ is the channel matrix at the $l$-th lag, $\vs[k]\in\mathbb{C}^{N_t}$ is the transmitted symbol, $\vw[k]$ is the noise, and the quantization operator $\mathcal{Q}$ quantizes the real and imaginary parts of a complex number respectively. The channel matrix $\mH[l]$ in the antenna domain can be expressed in terms of the angle domain representation $\mX[l]\in\mathbb{C}^{N_r\times N_t}$ as
\begin{align}
    \mH[l]=\mB_{N_r}\mX[l]\mB_{N_t}^*\,,
\end{align}
where $\mB_{N_r}\in\mathbb{C}^{N_r\times N_r}$ and $\mB_{N_t}\in\mathbb{C}^{N_t\times N_t}$ are the steering matrices for the receiver and transmitter arrays respectively, and they can be constructed from unitary discrete Fourier transform (DFT) matrices \cite{Sayeed:MIMOChannel:2002}. Rearranging the quantized measurements $y[k]$ from all the transmission blocks as in \cite{Mo:LowADC:2018}, we can simplify the notations of the quantized measurement model as follows
\begin{align}
    \vy=\mathcal{Q}(\mA\vx+\vw)\,,
\end{align}
where $\vy\in\mathbb{C}^M$ contains the quantized measurements, $\mA\in\mathbb{C}^{M\times N}$ is the measurement matrix, $\vx\in\mathbb{C}^N$ contains the (approximately) sparse channel coefficients in the angle domain, and $\vw\in\mathbb{C}^N$ contains the pre-quantization noises.

\begin{figure}[tbp]
\centering
\includegraphics[width=0.3\textwidth]{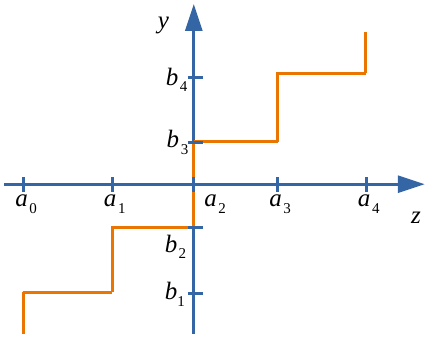}
\caption{The quantizer $\mathcal{Q}$ is applied on the real and imaginary parts of a complex measurement respectively.}
\label{fig:multi_bit_quantizer}
\end{figure}

Our goal is to recover the channel coefficient vector $\vx$ given $\vy$ and $\mA$. We shall use AMP to compute the posterior distribution $p(\vx|\vy)$ and the MMSE estimation $\hat{\vx}$
\begin{align}
    \hat{\vx}=\int\vx \cdot p(\vx|\vy) \ \mathrm{d}\vx\,.
\end{align}

\begin{signal_prior}
Here we assume the entries of the sparse signal $\vx$ are i.i.d. and follow the Bernoulli and Gaussian mixture distribution
\begin{align}
\label{eq:bgm}
    p(x_n) = (1-\kappa)\cdot\delta(x_n)+\kappa\cdot\sum_{i=1}^D\xi_i\cdot\mathcal{CN}(x_n|\mu_i,{\nu_x}_i)\,,
\end{align}
where $\delta(x_n)$ is the Dirac delta function, $\kappa$ is the probability that $x_n$ takes a non-zero value, $\xi_i$ is the Gaussian mixture weight, $\mu_i$ and ${\nu_x}_i$ are the mean and variance of the $i$-th complex Gaussian component, and $D$ is the number of complex Gaussian mixture components. We assume the pseudo-covariance (relation) of $x_n$ is insignificant. The probability density function $\mathcal{CN}(x_n|\mu_i,{\nu_x}_i)$ is then 
\begin{align}
\mathcal{CN}(x_n|\mu_i,{\nu_x}_i)=\frac{1}{\pi\cdot {\nu_x}_i}\exp\left(-\frac{|x_n-\mu_i|^2}{{\nu_x}_i}\right)\,.
\end{align}
\end{signal_prior}

\begin{noise_prior} The pre-quantization noise $\vw$ is modeled as circularly symmetric complex Gaussian with variance $\tau_w$. The probability density function of $w_m\in\mathbb{C}$ can be modeled as
\begin{align}
\label{eq:complex_gaussian_noise}
    p(w_m)=\frac{1}{\pi\cdot \tau_w}\exp\left(-\frac{|w_m|^2}{\tau_w}\right)\,.
\end{align}
Let $\vz=\mA\vx$ denote the noiseless measurement, $K$ denote the number of bits of the quantized measurement $y_m$, and $\{b_k\ |\ k=1,\cdots,2^K\}$ denote the quantization symbol. As shown in Fig. \ref{fig:multi_bit_quantizer}, the quantizer $\mathcal{Q}$ is applied on the real and imaginary parts of $z_m+w_m$ respectively.
\begin{itemize}
\item If the real coefficient $\mathrm{Re}(z_m+w_m)\in[a_{k-1},a_k)$, then
\begin{align}
    \mathrm{Re}(y_m)=&\mathcal{Q}\big(\mathrm{Re}(z_m+w_m)\big)=b_k\,,
\end{align}
where $a_{k-1}$ and $a_k$ are the lower and upper thresholds corresponding to the symbol $b_k$.
\item If the imaginary coefficient $\mathrm{Im}(z_m+w_m)\in[a_{l-1},a_l)$, then
\begin{align}
    \mathrm{Im}(y_m)=&\mathcal{Q}\big(\mathrm{Im}(z_m+w_m)\big)=b_l\,,
\end{align}
where $a_{l-1}$ and $a_l$ are the lower and upper thresholds corresponding to the symbol $b_l$.
\end{itemize}

\end{noise_prior}

\section{AMP with Built-in Parameter estimation}
\label{sec:amp_pe}

According to the distributions in \eqref{eq:bgm} and \eqref{eq:complex_gaussian_noise}, we have two sets of parameters in the signal prior and the noise prior respectively:
\begin{align}
\boldsymbol\lambda&=\left\{\left.\kappa,\xi_i,\mu_i,{\nu_x}_i\ \right|\ i=1,\cdots,D\right\}\\
\boldsymbol\theta &=\left\{\tau_w\right\}.
\end{align}

The computation involved in parameter estimation by previous AMP approaches becomes too difficult for the quantization noise model. As a result, they either use an approximated AWGN model or manually tune the noise distribution parameter. 

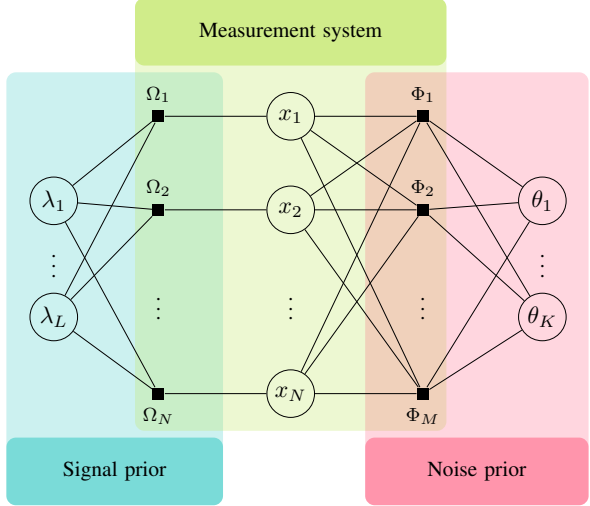
\begin{figure}[tbp]
\begin{center}
\begin{tabular}{c}
\hspace*{-1em}
\scalebox{0.9}{
%
%
%
%
\definecolor{00BCB4}{RGB}{0,188,180}
\definecolor{C4E86B}{RGB}{196,232,107}
\definecolor{FF3561}{RGB}{255,53,97}

\begin{tikzpicture}

  
  \fill[00BCB4, opacity=0.2, rounded corners] (-0.7,-4.5) rectangle (2.5,1.9);
  \fill[C4E86B, opacity=0.3, rounded corners] (1.2,-3.4) rectangle (5.8,3);
  \fill[FF3561, opacity=0.2, rounded corners] (4.6,-4.5) rectangle (7.9,1.9);

  \fill[00BCB4, opacity=0.4, rounded corners] (-0.7,-4.5) rectangle (2.5,-3.5);
  \node[text width=3cm, align=center] at (0.9,-4) {Signal prior};
  \fill[C4E86B, opacity=0.7, rounded corners] (1.2,2) rectangle (5.8,3);
  \node[text width=4.5cm, align=center] at (3.5,2.5) {Measurement system};
  \fill[FF3561, opacity=0.4, rounded corners] (4.6,-4.5) rectangle (7.9,-3.5);
  \node[text width=3cm, align=center] at (6.25,-4) {Noise prior};

  \node[latent,fill=00BCB4!20] (lambda_1) {$\lambda_1$};
  \node[latent,fill=00BCB4!20, below = 1 of lambda_1] (lambda_L) {$\lambda_L$};
  
  \path (lambda_1) -- node[auto=false]{\vdots} (lambda_L);

  \node[latent,fill=C4E86B!30, above=0.533 of lambda_1, xshift=3.5cm] (x_1) {$x_1$};
  \node[latent,fill=C4E86B!30, below=0.666 of x_1] (x_2) {$x_2$};
  \node[latent,fill=C4E86B!30, below=2 of x_2] (x_N) {$x_N$};
  
  \path (x_2) -- node[auto=false]{\vdots} (x_N);
  
  \factor[left=0 of x_1, xshift=-1.5cm] {Omega_1-f} {above:$\Omega_1$} {x_1, lambda_1, lambda_L}{};
  \factor[left=0 of x_2, xshift=-1.5cm] {Omega_2-f} {above:$\Omega_2$} {x_2, lambda_1, lambda_L}{};
  \factor[left=0 of x_N, xshift=-1.5cm] {Omega_N-f} {below:$\Omega_N$} {x_N, lambda_1, lambda_L}{};
  \path(Omega_2-f) -- node[auto=false]{\vdots} (Omega_N-f);
  
  \node[latent,fill=FF3561!20, right=0 of lambda_1, xshift=6.5cm] (theta_1) {$\theta_1$};
  \node[latent,fill=FF3561!20, right=0 of lambda_L, xshift=6.5cm] (theta_K) {$\theta_K$};
  \path(theta_1) -- node[auto=false]{\vdots} (theta_K);
  

  \factor[right=0 of x_1, xshift=1.5cm] {Phi_1-f} {above:$\Phi_1$} {theta_1, theta_K, x_1, x_2, x_N} {};
  \factor[right=0 of x_2, xshift=1.5cm] {Phi_2-f} {above:$\Phi_2$} {theta_1, theta_K, x_1, x_2, x_N} {};
  \factor[right=0 of x_N, xshift=1.5cm] {Phi_M-f} {below:$\Phi_M$} {theta_1, theta_K, x_1, x_2, x_N} {};
  
  \path(Phi_2-f) -- node[auto=false]{\vdots} (Phi_M-f);
\end{tikzpicture}

}
\end{tabular}
\end{center}
\vspace{-1em}
\caption{The factor graph of the sparse signal recovery task: ``$\bigcirc$'' represents the variable node, and ``$\blacksquare$'' represents the factor node \cite{Huang:1bitCS:2020}}
\vspace{-1em}
\label{fig:factor_graph_pegamp}
\end{figure}

As illustrated in Fig. \ref{fig:factor_graph_pegamp}, in this paper we choose the AMP framework introduced in \cite{PE_GAMP17} where the distribution parameters are treated as unknown random variables. The posteriors of the signal $\vx$ and the parameters $\{\boldsymbol\lambda,\boldsymbol\theta\}$ can then be jointly computed. Compared to previous parameter estimation approaches that maximize the likelihood \cite{Vila:EMGM:2013} or the Beth free entropy \cite{Krzakala:2012:1}, our proposed approach maximizes the posteriors of the parameters and could handle complicated distributions with ease.

We take advantage of the GAMP formulation \cite{Rangan:GAMP:2011} to compute the messages passed between the variable nodes and the factor nodes in Fig. \ref{fig:factor_graph_pegamp}. Taking the messages between the factor node $\Phi_m$ and the variable node $x_n$ for example, we use the following notations:
\begin{itemize}
\item $\Delta_{\Phi_m\rightarrow x_n}$ denotes the message from $\Phi_m$ to $x_n$,
\item $\Delta_{x_n\rightarrow \Phi_m}$ denotes the message from $x_n$ to $\Phi_m$.
\end{itemize}
Both $\Delta_{\Phi_m\rightarrow x_n}$ and $\Delta_{x_n\rightarrow \Phi_m}$ can be viewed as functions of $x_n$, and they are expressed in the ``$\log$'' domain. The messages are passed among the nodes iteratively until a consensus on how the variables should be distributed is reached by the factor nodes. 

In the $(t+1)$-th AMP iteration, the posteriors of the signal $\vx$ and the distribution parameters $\boldsymbol\lambda,\boldsymbol\theta$ can be obtained via belief propagation, aka sum-product message passing:
\begin{subequations}
\label{eq:sm_post_dist}
\begin{align}
\label{eq:pm_x}
\begin{split}
p(x_n|\vy)&\propto\exp\big(\Delta^{(t+1)}_{\Omega_n\rightarrow x_n}+\textstyle\sum_m\Delta^{(t+1)}_{\Phi_m\rightarrow x_n}\big)
\end{split}\\
\label{eq:pm_lambda}
\begin{split}
p(\lambda_l|\vy)&\propto\exp\big(\textstyle\sum_n\Delta^{(t+1)}_{\Omega_n\rightarrow\lambda_l}\big)
\end{split}\\
\label{eq:pm_theta}
\begin{split}
p(\theta_k|\vy)&\propto\exp\big(\textstyle\sum_m\Delta^{(t+1)}_{\Phi_m\rightarrow\theta_k}\big)\,.
\end{split}
\end{align}
\end{subequations}
Detailed derivations of the messages in \eqref{eq:sm_post_dist} can be found in our earlier work \cite{Huang:1bitCS:2020}. Note that the messages are derived for the real case in \cite{Huang:1bitCS:2020}, but the extension to the complex case for channel estimation should be straightforward. The distribution parameters $\boldsymbol\lambda,\boldsymbol\theta$ can be estimated by maximizing the posteriors in \eqref{eq:pm_lambda} and \eqref{eq:pm_theta}.
\begin{subequations}
\begin{align}
    \label{eq:lambda_est}
    \hat{\lambda}_l^{(t+1)} &= \arg\max_{\lambda_l}p(\lambda_l|\vy)=\arg\max_{\lambda_l}\textstyle\sum_n\Delta^{(t+1)}_{\Omega_n\rightarrow\lambda_l}\\
    \label{eq:theta_est}
    \hat{\theta}_k^{(t+1)} &= \arg\max_{\theta_k}p(\theta_k|\vy)=\arg\max_{\theta_k}\textstyle\sum_m\Delta^{(t+1)}_{\Phi_m\rightarrow\theta_k}\,.
\end{align}
\end{subequations}
We employ the computationally efficient approach in \cite{Huang:1bitCS:2020} to find the maximizing distribution parameters. It combines EM with the second-order method, and is much faster than vanilla gradient descent. 
When the measurement matrix $\mA$ is a zero-mean random Gaussian matrix, the convergence behavior of AMP in the large system limit as $N\rightarrow\infty$ can be characterized and guaranteed by state evolution \cite{Rangan:GAMP:2011}. It is still an open problem to establish convergence conditions of AMP for arbitrary matrices. Damping and mean-removal operations are often incorporated to help achieving convergence for the AMP algorithm \cite{Rangan:DampingCvg:2014,Vila:DampingMR:2015}.


\section{Experimental results}
\label{sec:experiments}


\begin{figure*}[htbp]
\centering
\subfigure{
\label{fig:channel_1}
\includegraphics[width=0.9\textwidth]{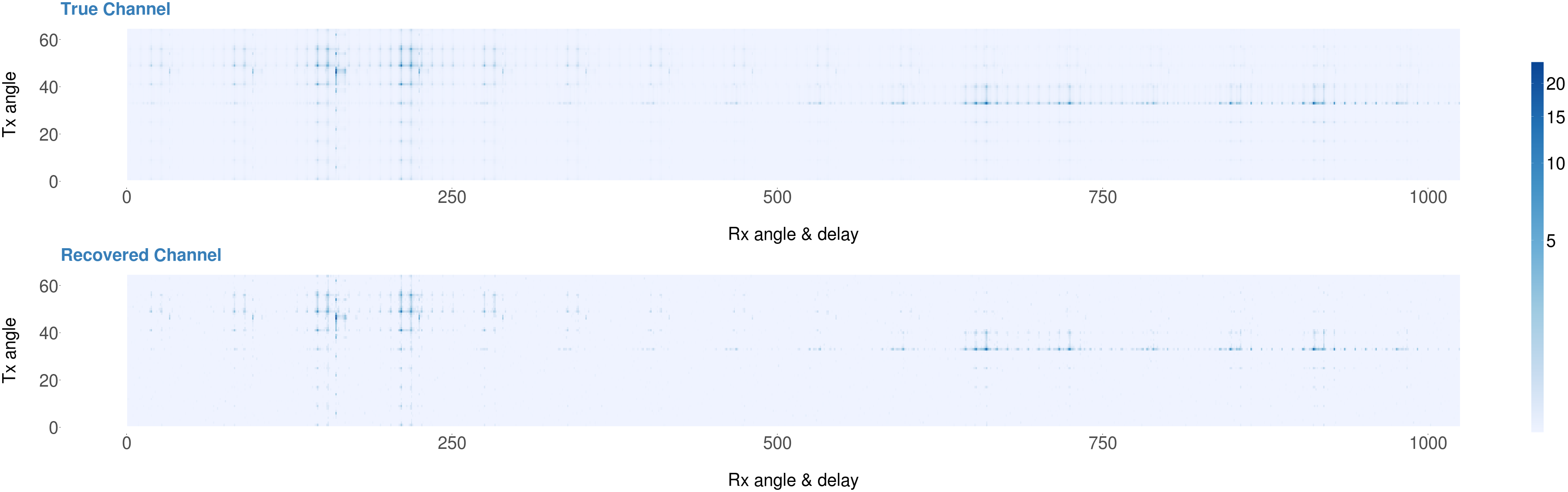}}
\vspace{-1em}
\caption{The magnitudes (in log scale) of the true channel $\vx$ and the recovered channel $\hat{\vx}$ from 1-bit measurements using the proposed AMP-PE approach when SNR$=10$ dB. The NMSE of the recovered channel $\hat{\vx}$ is $-12.02$ dB. }
\vspace{-1em}
\label{fig:channel_compare}
\end{figure*}

We next compare our proposed AMP with built-in parameter estimation framework (AMP-PE) with other state-of-the-art approaches such as the AMP that uses an approximated AWGN model (AMP-AWGN) \cite{Rangan:GAMP:2011,Vila:EMGM:2013}, least square approach (LS), the iterative hard thresholding (IHT) \cite{Blumensath2008}, and basis pursuit (BP) \cite{Chen:BP:1998}. In particular, IHT requires the knowledge of the sparsity of the signal, BP requires the knowledge of the noise level. These parameters need to be manually tuned in practice. Reproducible code and data are available at \urlstyle{tt}\url{https://github.com/shuai-huang/Channel-Est}

We estimate a broadband mmWave massive MIMO channel that has $N_{cl}=4$ clusters (each has $10$ paths) and a delay spread of at most $L=16$ symbols in the single-user case. Here we consider a practical setting in \cite{Chataut:mMIMO} where the transmitter and receiver are both equipped with $8\times8=64$ UPAs, corresponding to $N_t=64$ antennas at the transmitter and $N_r=64$ antennas at the receiver. The length of the channel coefficient vector $\vx$ is then $N=N_tN_rL$. A random QPSK block transmission of length $N_p=2048$ is used for training \cite{Mezghani:QPSK:2007}, producing $M=N_rN_p$ quantized measurements in $\vy$ with an oversampling rate $\frac{M}{N}=2$. 

Let $\vr=\mA\vx+\vw$ denote the unquantized measurements. The signal-to-noise ratio (SNR) of $\vr$ varies between $0$ and $40$ dB in the experiments, ranging from high-noise to low-noise regimes. Normalized mean squared error (NMSE) of the recovered signal $\hat{\vx}$ is computed and used to evaluate the performances of all approaches.
\begin{align}
\textnormal{NMSE}(\hat{\vx})=10\log_{10}\left(\mathbb{E}\left[\textstyle\frac{\|\hat{\vx}-\vx\|_2^2}{\|\vx\|_2^2}\right]\right)\,.
\end{align}

Fig. \ref{fig:channel_compare} illustrates a comparison of the true channel $\vx$ and the recovered channel $\hat{\vx}$ from 1-bit measurements using the proposed AMP-PE approach when SNR$=10$ dB. We are able to achieve a decent recovery with NMSE$=-12.02$ dB from only 1-bit measurements. The performance can be improved further with more measurements or higher-resolution ADCs. 

\begin{figure*}[h]
\centering
\subfigure{
\label{fig:1bit}
\includegraphics[width=0.25\textwidth]{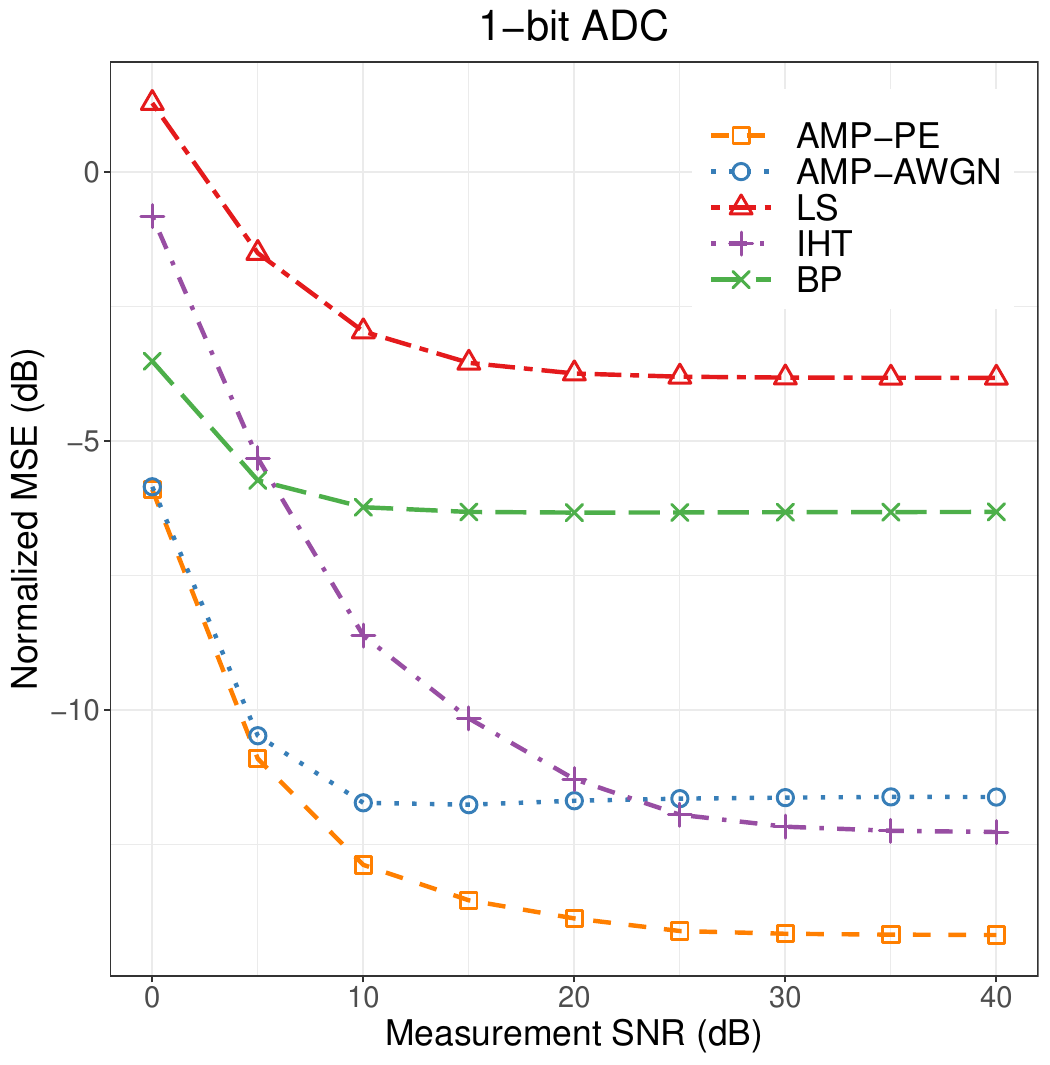}}
\subfigure{
\label{fig:2bit}
\includegraphics[width=0.25\textwidth]{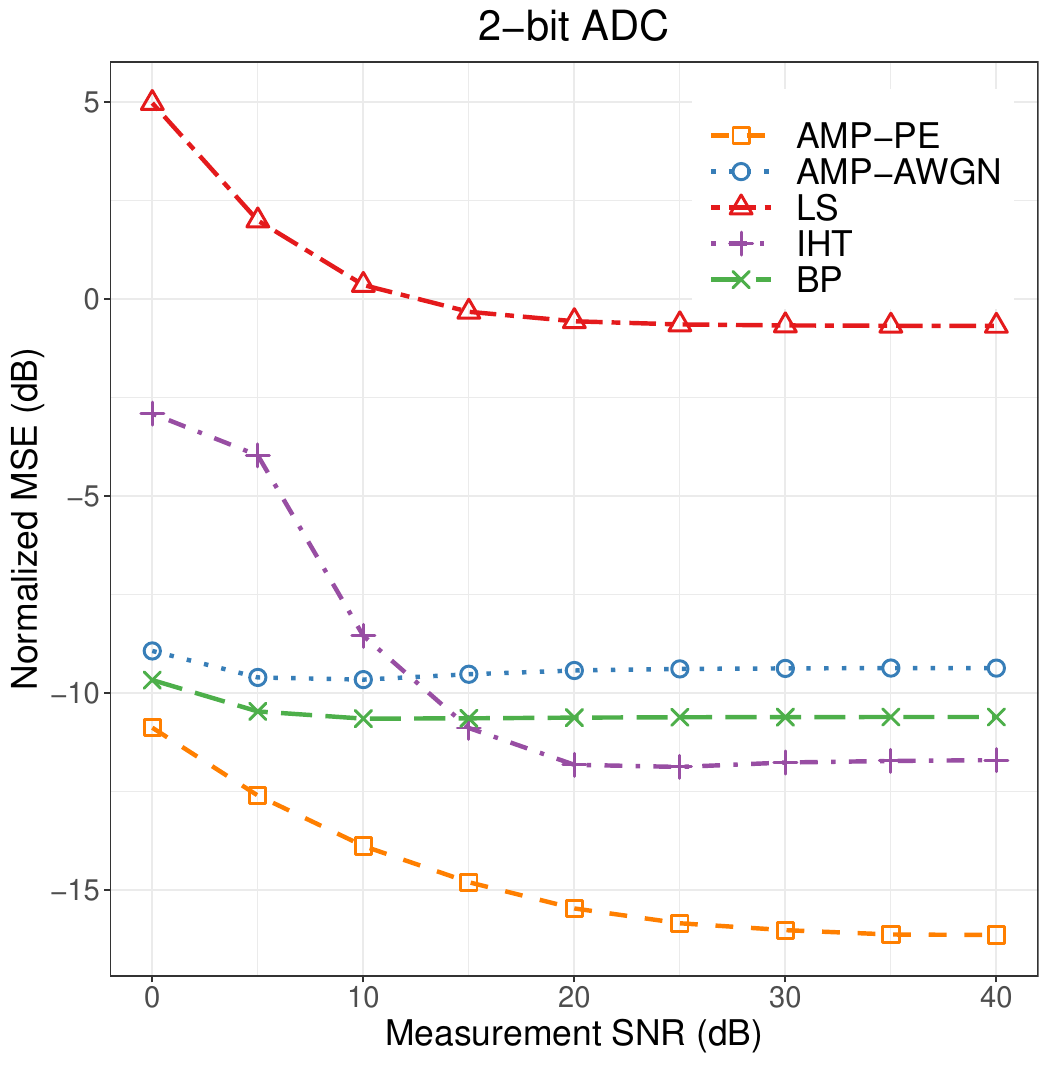}}
\subfigure{
\label{fig:3bit}
\includegraphics[width=0.25\textwidth]{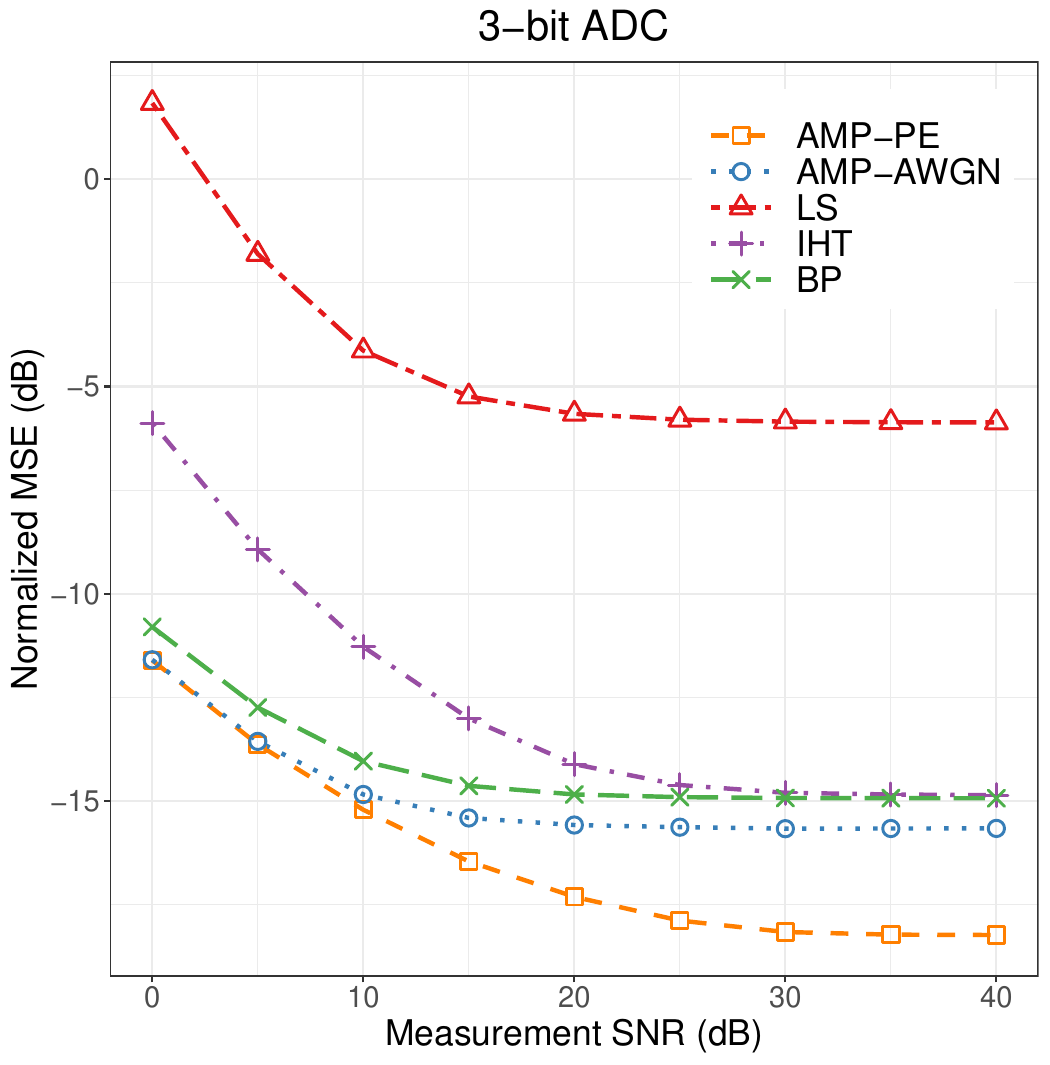}}
\vspace{-1em}
\caption{Comparison of the proposed AMP-PE approach with the AMP-AWGN approach, the least square (LS), iterative hard thresholding (IHT) and basis pursuit (BP) across different noise levels. The quantized measurements are from 1-bit, 2-bit and 3-bit ADCs.}
\vspace{-1em}
\label{fig:aproach_compare}
\end{figure*}

The average NMSEs from 100 random trials using different approaches are depicted in Fig. \ref{fig:aproach_compare}. The proposed AMP-PE adopts the quantization noise model and automatically estimates the distribution parameters by maximizing their posteriors. The AMP-AWGN relies on an approximated AWGN model and computes the maximum likelihood estimations of the distribution parameters via EM. Damping is applied on the recovered $\hat{\vx}$ at a rate of $0.1$ to stabilize the convergence of both AMP-PE and AMP-AWGN. The parameters of IHT and BP are manually tuned to achieve their best performances. For 1-bit recovery, we can see that AMP-PE performs much better than other approaches in the moderate-noise and low-noise regimes (SNR$\geq10$ dB), and slightly better than AMP-AWGN in the high-noise regime (SNR$<10$ dB). For 2-bit recovery, AMP-PE performs significantly better than other approaches in all cases. For 3-bit recovery, AMP-PE still outperforms the other approaches in the moderate-noise and low-noise regimes (SNR$\geq10$ dB), while performing almost equally well as AMP-AWGN in the high-noise regime (SNR$<10$ dB). In general, the proposed AMP-PE approach shows a significant advantage over other state-of-the-art methods. It offers a more practical and maintenance-free alternative where the signal and parameters can be jointly recovered based only on inaccurate snapshots of the environments. 

\section{Conclusion}
\label{sec:conlucsion}
Assuming only sparse prior of the massive MIMO channel coefficients $\vx$ in the angle domain, we perform channel estimation from a Bayesian perspective where the signal $\vx$ and the quantized measurements $\vy$ follow certain distributions with unknown parameters. We propose to jointly recover the signal and parameters within the extended AMP framework where they are both treated as random variables. Due to the nonlinearity of the quantization noise, previous AMP-based channel estimation methods either use an approximated noise model or manually tune the noise distribution parameters. Our proposed approach works with the actual quantization noise model and can estimate model parameters by maximizing their posteriors.

For non-Gaussian measurement matrices in channel estimation, we need to apply damping on the recovered $\hat{\vx}$ to stabilize the convergence. In the experiments we discover that the initialization of parameters also affects the algorithm's convergence behavior. We plan to further investigate which role the initialization plays and to design an efficient initialization strategy in the future work. Notwithstanding these drawbacks, the proposed framework achieves state-of-the-art performances. We also believe that the framework enjoys a wider applicability in practice compared to methods that require parameter tuning for different types of signals and noise levels.


\bibliographystyle{IEEEbib}
\bibliography{refs}

\begin{thebibliography}{10}

\bibitem{Larsson:massiveMIMO:2014}
E.~G. {Larsson}, O.~{Edfors}, F.~{Tufvesson}, and T.~L. {Marzetta},
\newblock ``Massive mimo for next generation wireless systems,''
\newblock {\em IEEE Communications Magazine}, vol. 52, no. 2, pp. 186--195,
  2014.

\bibitem{Lu:massiveMIMO:2014}
L.~{Lu}, G.~Y. {Li}, A.~L. {Swindlehurst}, A.~{Ashikhmin}, and R.~{Zhang},
\newblock ``An overview of massive mimo: Benefits and challenges,''
\newblock {\em IEEE Journal of Selected Topics in Signal Processing}, vol. 8,
  no. 5, pp. 742--758, 2014.

\bibitem{Asaad:massiveMIM0:2018}
S.~{Asaad}, A.~M. {Rabiei}, and R.~R. {Müller},
\newblock ``Massive mimo with antenna selection: Fundamental limits and
  applications,''
\newblock {\em IEEE Trans. Wirel. Commun.}, vol. 17, no. 12, pp. 8502--8516,
  2018.

\bibitem{Walden:ADC:1999}
R.~H. {Walden},
\newblock ``Analog-to-digital converter survey and analysis,''
\newblock {\em IEEE Journal on Selected Areas in Communications}, vol. 17, no.
  4, pp. 539--550, 1999.

\bibitem{Murmann:1997-2020}
B.~Murmann,
\newblock {\em ADC Performance Survey 1997-2020},
\newblock [Online]. Available:
  \url{http://web.stanford.edu/~murmann/adcsurvey.html}.

\bibitem{Rappaport:MmWave:2015}
T.~Rappaport, R.~Heath, R.~Daniels, and J.~Murdock,
\newblock {\em Millimeter wave wireless communications},
\newblock Prentice Hall, 2015.

\bibitem{Dabeer:CEADC:2010}
O.~{Dabeer} and U.~{Madhow},
\newblock ``Channel estimation with low-precision analog-to-digital
  conversion,''
\newblock in {\em 2010 IEEE International Conference on Communications}, 2010,
  pp. 1--6.

\bibitem{Zeitler:DitherCE:2012}
G.~{Zeitler}, G.~{Kramer}, and A.~C. {Singer},
\newblock ``Bayesian parameter estimation using single-bit dithered
  quantization,''
\newblock {\em IEEE Trans. Signal Process.}, vol. 60, no. 6, pp. 2713--2726,
  2012.

\bibitem{Sayeed:MIMOChannel:2002}
A.~M. {Sayeed},
\newblock ``Deconstructing multiantenna fading channels,''
\newblock {\em IEEE Trans. Signal Process.}, vol. 50, no. 10, pp. 2563--2579,
  2002.

\bibitem{Rao:CESparse:2014}
X.~{Rao} and V.~K.~N. {Lau},
\newblock ``Distributed compressive csit estimation and feedback for fdd
  multi-user massive mimo systems,''
\newblock {\em IEEE Trans. Signal Process.}, vol. 62, no. 12, pp. 3261--3271,
  2014.

\bibitem{5G:Report:2016}
``5g channel model for bands up to 100 ghz (v2.3), aalto university, at\&t,
  bupt, cmcc, ericsson, huawei, intel, kt corporation, nokia, ntt docomo, new
  york university, qualcomm, samsung, university of bristol, university of
  southern california,''
\newblock pp. 1--107, October 2016,
\newblock [Online]. Available: \url{http://www.5gworkshops.com/5gcm.html}.

\bibitem{Decode05}
E.~J. Cand{\`e}s and T.~Tao,
\newblock ``Decoding by linear programming,''
\newblock {\em IEEE Trans. Inf. Theory}, vol. 51(12), pp. 4203--4215, 2005.

\bibitem{CS06}
D.~L. Donoho,
\newblock ``Compressed sensing,''
\newblock {\em IEEE Trans. Inf. Theory}, vol. 52, no. 4, pp. 1289--1306, 2006.

\bibitem{Blumensath2008}
T.~Blumensath and M.~E. Davies,
\newblock ``Iterative thresholding for sparse approximations,''
\newblock {\em Journal of Fourier Analysis and Applications}, vol. 14, no. 5,
  pp. 629--654, Dec 2008.

\bibitem{Jacques:1bitCS:2013}
L.~{Jacques}, J.~N. {Laska}, P.~T. {Boufounos}, and R.~G. {Baraniuk},
\newblock ``Robust 1-bit compressive sensing via binary stable embeddings of
  sparse vectors,''
\newblock {\em IEEE Trans. Inf. Theory}, vol. 59, no. 4, pp. 2082--2102, 2013.

\bibitem{l1stable06}
E.~J. Cand{\`e}s, J.~K. Romberg, and T.~Tao,
\newblock ``Stable signal recovery from incomplete and inaccurate
  measurements,''
\newblock {\em Commun. Pure Appl. Math.}, vol. 59, no. 8, pp. 1207--1223, 2006.

\bibitem{Nonconvex_lp07}
R.~Chartrand,
\newblock ``Exact reconstructions of sparse signals via nonconvex
  minimization,''
\newblock {\em IEEE Signal Processing Letters}, vol. 14, pp. 707--710, 2007.

\bibitem{Huang:Entropy:2019}
S.~{Huang} and T.~D. {Tran},
\newblock ``Sparse signal recovery via generalized entropy functions
  minimization,''
\newblock {\em IEEE Trans. Signal Process.}, vol. 67, no. 5, pp. 1322--1337,
  2019.

\bibitem{Minka:2001}
T.~P. Minka and R.~Picard,
\newblock {\em A Family of Algorithms for Approximate Bayesian Inference},
\newblock Ph.D. thesis, USA, 2001.

\bibitem{Donoho:AMP:2009}
D.~L. Donoho, A.~Maleki, and A.~Montanari,
\newblock ``Message-passing algorithms for compressed sensing,''
\newblock {\em PNAS}, vol. 106, no. 45, pp. 18914--18919, 2009.

\bibitem{Metzler:Denoising:2016}
C.~A. {Metzler}, A.~{Maleki}, and R.~G. {Baraniuk},
\newblock ``From denoising to compressed sensing,''
\newblock {\em IEEE Trans. Inf. Theory}, vol. 62, no. 9, pp. 5117--5144, Sep.
  2016.

\bibitem{Ma:AMP_Denoise:2016}
Y.~{Ma}, J.~{Zhu}, and D.~{Baron},
\newblock ``Approximate message passing algorithm with universal denoising and
  gaussian mixture learning,''
\newblock {\em IEEE Trans. on Signal Process.}, vol. 64, no. 21, pp.
  5611--5622, 2016.

\bibitem{Rangan:GAMP:2011}
S.~Rangan,
\newblock ``Generalized approximate message passing for estimation with random
  linear mixing,''
\newblock in {\em Proceedings of IEEE ISIT}, July 2011, pp. 2168--2172.

\bibitem{Krzakala:2012:1}
F.~Krzakala, M.~M\'ezard, F.~Sausset, Y.~F. Sun, and L.~Zdeborov\'a,
\newblock ``Statistical-physics-based reconstruction in compressed sensing,''
\newblock {\em Phys. Rev. X}, vol. 2, pp. 021005, May 2012.

\bibitem{Wen:LowADC:2016}
C.~{Wen}, C.~{Wang}, S.~{Jin}, K.~{Wong}, and P.~{Ting},
\newblock ``Bayes-optimal joint channel-and-data estimation for massive mimo
  with low-precision adcs,''
\newblock {\em IEEE Trans. Signal Process.}, vol. 64, no. 10, pp. 2541--2556,
  2016.

\bibitem{Mo:LowADC:2018}
J.~{Mo}, P.~{Schniter}, and R.~W. {Heath},
\newblock ``Channel estimation in broadband millimeter wave mimo systems with
  few-bit adcs,''
\newblock {\em IEEE Trans. Signal Process.}, vol. 66, no. 5, pp. 1141--1154,
  2018.

\bibitem{Bellili:Lap:2019}
F.~{Bellili}, F.~{Sohrabi}, and W.~{Yu},
\newblock ``Generalized approximate message passing for massive mimo mmwave
  channel estimation with laplacian prior,''
\newblock {\em IEEE Transactions on Communications}, vol. 67, no. 5, pp.
  3205--3219, 2019.

\bibitem{Dempster:EM:1977}
A.~P. Dempster, N.~M. Laird, and D.~B. Rubin,
\newblock ``Maximum likelihood from incomplete data via the em algorithm,''
\newblock {\em J. R. Stat. Soc., Series B}, vol. 39, no. 1, pp. 1--38, 1977.

\bibitem{Vila:EMGM:2013}
J.~P. {Vila} and P.~{Schniter},
\newblock ``Expectation-maximization gaussian-mixture approximate message
  passing,''
\newblock {\em IEEE Trans. Signal Process.}, vol. 61, no. 19, pp. 4658--4672,
  2013.

\bibitem{risi2014massive}
C.~Risi, D.~Persson, and E.~G. Larsson,
\newblock ``Massive mimo with 1-bit adc,''
\newblock {\em arXiv:1404.7736}, 2014.

\bibitem{Wang:Multiuser:2015}
S.~{Wang}, Y.~{Li}, and J.~{Wang},
\newblock ``Multiuser detection in massive spatial modulation mimo with
  low-resolution adcs,''
\newblock {\em IEEE Trans. Wirel. Commun.}, vol. 14, no. 4, pp. 2156--2168,
  2015.

\bibitem{Wen:Channel:2015}
C.~{Wen}, S.~{Jin}, K.~{Wong}, J.~{Chen}, and P.~{Ting},
\newblock ``Channel estimation for massive mimo using gaussian-mixture bayesian
  learning,''
\newblock {\em IEEE Trans. Wirel. Commun.}, vol. 14, no. 3, pp. 1356--1368,
  March 2015.

\bibitem{PE_GAMP17}
S.~Huang and T.~D. Tran,
\newblock ``Sparse signal recovery using generalized approximate message
  passing with built-in parameter estimation,''
\newblock in {\em Proceedings of IEEE ICASSP}, March 2017, pp. 4321--4325.

\bibitem{Huang:1bitCS:2020}
S.~Huang, D.~Qiu, and T.~D. Tran,
\newblock ``1-bit compressive sensing via approximate message passing with
  built-in parameter estimation,''
\newblock {\em arXiv:2007.07679}, 2020.

\bibitem{Rangan:DampingCvg:2014}
S.~{Rangan}, P.~{Schniter}, and A.~{Fletcher},
\newblock ``On the convergence of approximate message passing with arbitrary
  matrices,''
\newblock in {\em Proceedings of IEEE ISIT}, 2014, pp. 236--240.

\bibitem{Vila:DampingMR:2015}
J.~{Vila}, P.~{Schniter}, S.~{Rangan}, F.~{Krzakala}, and L.~{Zdeborová},
\newblock ``Adaptive damping and mean removal for the generalized approximate
  message passing algorithm,''
\newblock in {\em Proceedings of IEEE ICASSP}, 2015, pp. 2021--2025.

\bibitem{Chen:BP:1998}
S.~S. Chen, D.~L. Donoho, and M.~A. Saunders,
\newblock ``Atomic decomposition by basis pursuit,''
\newblock {\em SIAM Journal on Scientific Computing}, vol. 20, no. 1, pp.
  33--61, 1998.

\bibitem{Chataut:mMIMO}
R.~Chataut and R.~Akl,
\newblock ``Massive mimo systems for 5g and beyond networks—overview, recent
  trends, challenges, and future research direction,''
\newblock {\em Sensors}, vol. 20, no. 10, 2020.

\bibitem{Mezghani:QPSK:2007}
A.~{Mezghani} and J.~A. {Nossek},
\newblock ``On ultra-wideband mimo systems with 1-bit quantized outputs:
  Performance analysis and input optimization,''
\newblock in {\em Proceedings of IEEE ISIT}, 2007, pp. 1286--1289.

\end{thebibliography}

\end{document}